\documentclass[journal,final]{IEEEtran}
\usepackage{cite}
\ifCLASSINFOpdf
  \usepackage[pdftex]{graphicx}
\else
  \usepackage[dvips]{graphicx}
\fi

\usepackage{color,soul}
\sethlcolor{yellow} 
\usepackage[cmex10]{amsmath}
\usepackage{siunitx}
\usepackage{amssymb}
\usepackage{bm}
\usepackage{textcomp}
\usepackage{array}
\usepackage{dblfloatfix}
\usepackage{url}
\begin{document}
\bstctlcite{IEEEexample:BSTcontrol}
%
% paper title
% Titles are generally capitalized except for words such as a, an, and, as,
% at, but, by, for, in, nor, of, on, or, the, to and up, which are usually
% not capitalized unless they are the first or last word of the title.
% Linebreaks \\ can be used within to get better formatting as desired.
% Do not put math or special symbols in the title.
\title{Estimation of the electrostatic effects in the LISA-Pathfinder critical test mass dynamics via the method of moments}
%
%
% author names and IEEE memberships
% note positions of commas and nonbreaking spaces ( ~ ) LaTeX will not break
% a structure at a ~ so this keeps an author's name from being broken across
% two lines.
% use \thanks{} to gain access to the first footnote area
% a separate \thanks must be used for each paragraph as LaTeX2e's \thanks
% was not built to handle multiple paragraphs
%

\author{Carlo~Zanoni and Daniele~Bortoluzzi
\thanks{C. Zanoni is with the Directorate of Engineering, European Southern Observatory (ESO), Munich,  Germany e-mail: carlo.zanoni@alumni.unitn.it}% <-this % stops a space
\thanks{D. Bortoluzzi is with the Department of Industrial Engineering, University of Trento, Trento,  Italy, and also with Trento Institute of Fundamental Physics and Applications, INFN, Trento, Italy.}}% <-this % stops a space
\maketitle

% As a general rule, do not put math, special symbols or citations
% in the abstract or keywords.
\begin{abstract}
LISA-Pathfinder is an ESA space mission flown between 2015 and 2017 to demonstrate a technological maturity sufficient for building a gravitational waves telescope in space, such as the Laser Interferometer Space Antenna (LISA). A pair of cubic test masses is hosted inside the LISA-Pathfinder spacecraft and shielded from any force other than the interplanetary gravitational field. The purity of the shielding gives the performance of the mission.

There are a number of aspects that had to be confirmed in-flight. One of them is the transition phase from the launch configuration, when the test masses are locked, to the science free-falling configuration. Each test mass is initially released from the mechanical constraints via a dedicated mechanism and then captured by an electrostatic control system. In fact, each test mass is surrounded by a set of electrodes for actuation and sensing purposes. The performance criterion of the release is the final velocity of the test mass relative to the spacecraft, with an upper threshold set to 5~\si{\micro \metre / \second}. The LISA-Pathfinder first in-flight release velocities highlighted an unexpected dynamics with large linear and angular velocities. The electrostatic control was successful, but only relying on a manual procedure that cannot be considered as baseline for LISA.

This paper helps investigating the in-flight non-compliance by dealing with the modeling of the electrostatic environment around each test mass and its contribution to the release and capture dynamics. The electrostatic model is based on the method of moments, a boundary element numerical technique suitable for estimating forces and capacitances between conductors. We also provide a short overview of the method, which can be used for the analysis of other phenomena within LISA and for the design of future gravitational waves telescopes and space projects. 
\end{abstract}

% Note that keywords are not normally used for peerreview papers.

% For peer review papers, you can put extra information on the cover
% page as needed:
% \ifCLASSOPTIONpeerreview
% \begin{center} \bfseries EDICS Category: 3-BBND \end{center}
% \fi
%
% For peerreview papers, this IEEEtran command inserts a page break and
% creates the second title. It will be ignored for other modes.
\IEEEpeerreviewmaketitle

\section{Introduction}

\IEEEPARstart{T}{he} LISA-Pathfinder (LPF) space mission \cite{Armano2016} is an European Space Agency (ESA) project launched on December~3,~2015 and in operation until July 2017. The purpose of the mission was the experimental demonstration of a free fall of cubic test masses (TMs) suitable for a gravitational waves space observatory, such as the Laser Interferometer Space Antenna (LISA) \cite{proposal}. LISA will be an L-class ESA mission and it will benefit from the lessons learned with the development and operation of LPF.

\begin{center}
\begin{figure}[h!]
\includegraphics[width=\columnwidth]{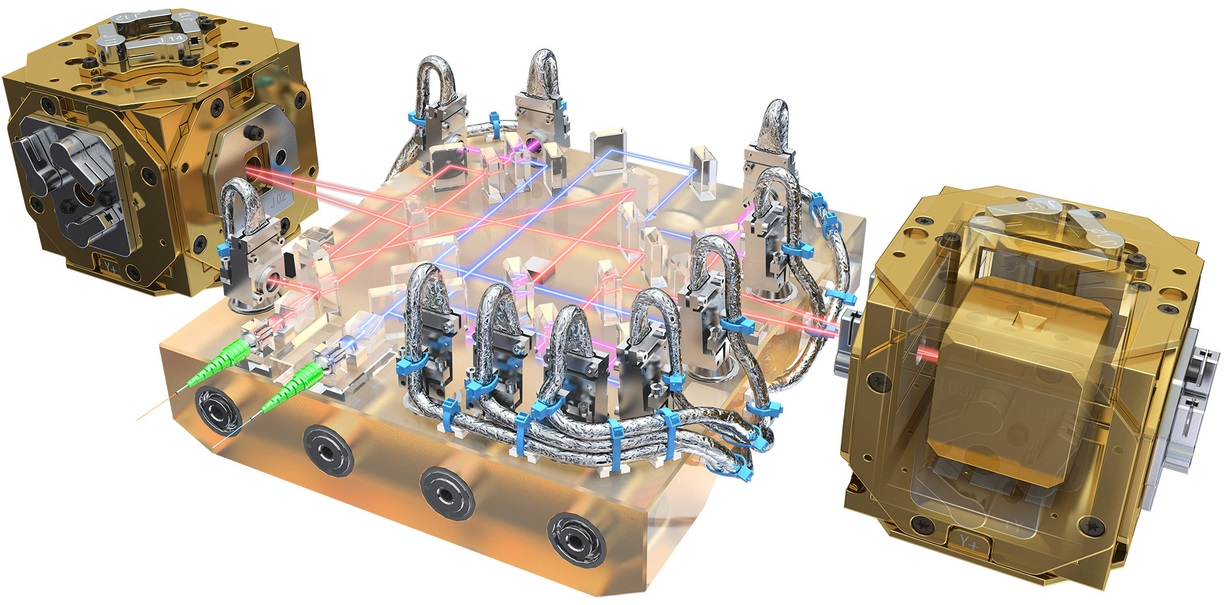}
\caption{Artist's impression of the LISA Technology Package core assembly, courtesy of ESA/ATG medialab.}
\label{fig0}
\end{figure}
\end{center}

The working principle of LPF sees two identical free falling TMs whose relative distance is measured by an interferometric optical system. Each TM is hosted in an Electrode Housing (EH) capable of actuation and sensing, Fig.~\ref{fig0}. The condition of free fall, or drag free \cite{Lange}, requires that the effects of every force but interplanetary gravity are minimized. This includes obviously complete absence of any mechanical constraint. The main performance measurement of LPF is given by the Amplitude Spectral Density (ASD) of the TMs differential acceleration noise within the frequency band 1~\si{\milli \hertz} $\leq f \leq$ 30~\si{\milli \hertz}. The final results \cite{LPF2} show an ASD well below the target and already in line with the expected LISA requirements.

Several noise sources are demonstrated being dependent on the TM-to-EH gaps \cite{John}, which for this reason are in the 2.9~$-$~4~\si{\milli \metre} range \cite{LPFcharge1}. Such large gaps means that, if the TM was left unconstrained during launch, the collision forces would not be acceptable \cite{ESAobjectinjection}. A dedicated lock and release mechanism is therefore deemed necessary. Besides, the large TM-to-EH gap is limiting the EH electrodes authority. The requirements on the lock and release mechanism are therefore tight, to allow the EH actuation to initialize the science phase. The maximum velocity of the TM right after the separation from the residual mechanical constraints is 5~\si{\micro \metre / \second} and 100~\si{\micro \radian/\second} about any linear and angular axis. Besides, according to the nominal configuration, the residual velocity is expected to be mostly along the linear axis of retraction of the two opposed release-dedicated tips, Fig.~\ref{fig0}.

The lock and release mechanism was one of the critical technologies to be validated by LPF. The literature covering the on-ground estimations, with a focus on the adhesion contribution \cite{MSSP13,Trans}, indicates a TM velocity well within the requirement.  However, the in-flight results showed release velocities as high as 56.7~\si{\micro \metre / \second} revealing an unexpected dynamics. Nonetheless, it was eventually possible to electrostatically capture the TMs every time it was required, by commanding manually the mechanism \cite{Bortoluzzi2020,esmats19} and thanks to the control system performance. 

The manual solution found for LPF is not an option for LISA due to the number of TMs and the distance of the three spacecraft from the Earth. What happened in LPF needs then to be understood to deliver a more robust design for LISA. A post-flight analysis of mechanical effects is presented in \cite{Bortoluzzi2020,esmats19}. This paper deals instead with the possible electrostatic contributions.
%: the force between a charged TM and the release mechanism, and the un-modeled biases in the TM position and velocity measurements, which is based on capacitance sensing. 

The electrostatic effects are modeled via the Method of Moments (MoM) \cite{MOM1}, a Boundary Element Method. The references in literature, such as \cite{MOM2}, often focus on the use of the MoM for radiation and scattering problems. In the first section of this paper, we therefore overview the MoM in the scope of a force estimation between isolated charged solid bodies. This summary is intended to maximize reproducibility of the results and of the modeling technique, which is of interest for the analysis of other aspects of LISA where the capacitance matrix, the charge distribution or the electric field matter. The second section introduces the reader to technical aspects of the LPF gravitational reference sensor (GRS). The last section presents the estimation of the effects analyzed, compared to the release requirements.

%%%%%%%%%%%%%%%%%%%%%%%%%%%%%%%%%%%%%%%%%%%%%%%%%%%%%%%%%%%%%%%%%%%%%%%%%%%%%%%%%%%%%%%%%%%%%%%%%%%%%%%%%%%%%%%%%%%%%%%%%%%%%%%%%%%%%%%%%%%%%%%%%%%%%%%%%%%%%%%%
\section{The Method of Moments applied to solid charged bodies}
\label{sec1}

The MoM is a numerical technique in which only the surfaces of the conductors are discretized. It is preferred to the Finite Element Method because it is more easily implemented and customized with a programming code. Being a Boundary Element Method, the MoM is also more efficient for problems with a small surface-to-volume ratio and encounters fewer issues in modeling the \si{\micro \metre} gaps encountered in the LPF release.

Let's start the overview recalling the voltage produced by a charged body of volume $V$:
\begin{equation}
	\phi(\textbf{r}) = \int_{V} \frac{\sigma(\textbf{r}^*)}{4\pi\epsilon_0 \left| \textbf{r} - \textbf{r}^* \right|} d\textbf{r}^*
\label{volt1}
\end{equation}
where $\epsilon_0$ is the dielectric constant, $\phi(\textbf{r})$ is the voltage at a point $\textbf{r}$ and $\sigma(\textbf{r}^*)$ is a function describing the charge density in $\mathbb{R}^3$.

\begin{center}
\begin{figure}[h]
\includegraphics[width=0.8\columnwidth]{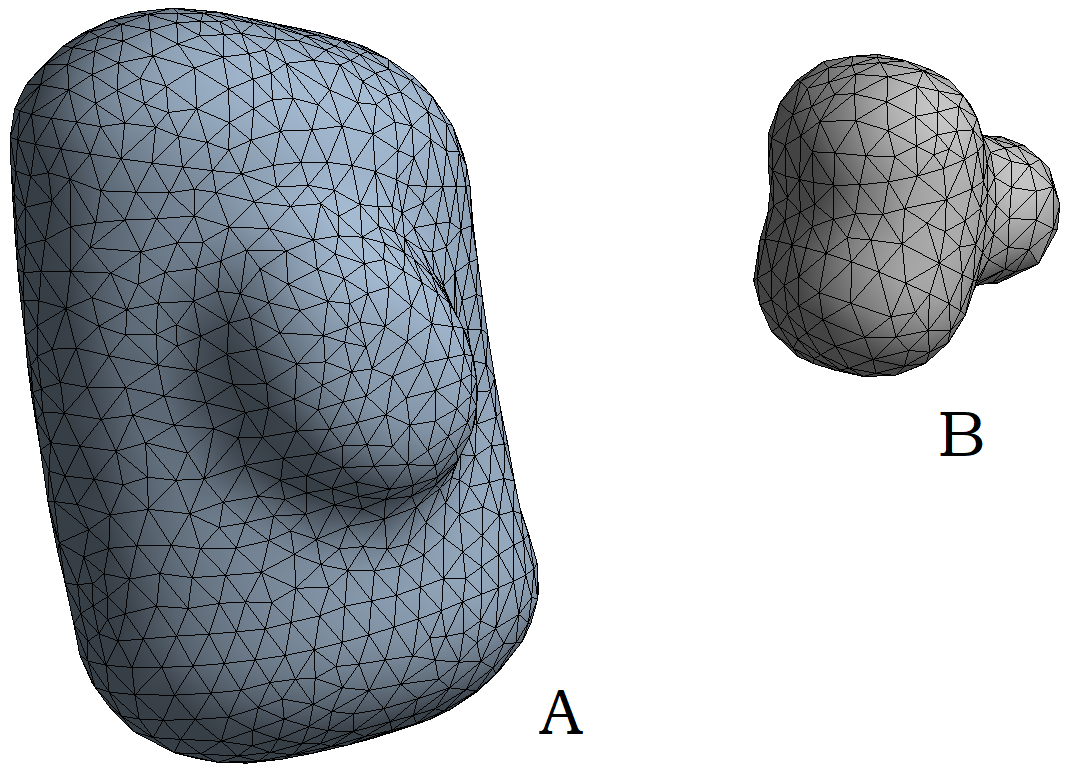}
\caption{Two conductors of generic shape discretized with a triangular mesh. The method of moments aims at finding the relationship between the vector of voltages of all the triangles and the vector of charges. The use of conductors simplifies the boundary conditions because the voltage can be assumed uniform in all the mesh elements belonging to the same object.}
\label{fig1}
\end{figure}
\end{center}

To help visualizing the problem one can consider the conductors of Fig.~\ref{fig1} for this recap.
By definition, all the charges are on the conductors surfaces, which means that the volume integrals can be transformed in surface integrals. The surfaces are meshed with triangles such that body A has $N_A$ elements and body B has $N_B$. The triangular mesh is here chosen because it allows an easier discretization of complex shapes. For a triangle with index $n$ the charge density is here assumed constant with value $\sigma_n$.

The voltage due to a mesh element $n$, with surface $S_n$, is given by:
\begin{equation}
	\phi_n(\textbf{r}) = \int_{S_n} \frac{\sigma_n}{4\pi\epsilon_0 \left| \textbf{r} - \textbf{r}^* \right|} d\textbf{r}^*
\label{volt2}
\end{equation}
If such a voltage is then calculated and integrated on the surface of an element $m$ the following is obtained:
\begin{equation}
	\int_{S_m} \phi_n(\textbf{r}) d\textbf{r} = \left(\int_{S_m} \int_{S_n} \frac{1}{4\pi\epsilon_0 \left| \textbf{r} - \textbf{r}^* \right|} d\textbf{r}^* d\textbf{r} \right) \sigma_n = Z_{mn} \sigma_n %\int_{S_m} \int_{S_n} \frac{f_n(\textbf{r}^*) f_m(\textbf{r}) }{4\pi\epsilon_0 \left| \textbf{r} - \textbf{r}^* \right|} d\textbf{r}^* d\textbf{r}
\label{volt3}
\end{equation}
%$Z_{mn}$ expresses the product of the surface $S_m$ and the voltage produced by the element $n$ on the element $m$, when a unitary charge density is present in the element $n$.
where a new parameter, $Z_{mn}$, is introduced. Except for physical constants, $Z_{mn}$ depends only on the position and the geometrical properties of the mesh elements of the conductors.

The total voltage of an element $m$ is given by the sum of the contributions of all the mesh triangles:
\begin{equation}
	\Phi_m = \sum_{n} \phi_n
\label{volt4}
\end{equation}

Eq.~\ref{volt3}, via eq.~\ref{volt4}, then becomes:
\begin{equation}
	S_m \Phi_m = \sum_{n} (Z_{mn} \sigma_n)
\label{volt5}
\end{equation}
and can be written as the following linear system:
\begin{equation}
\left[S\right] \left\{\Phi\right\} = \left[Z\right] \left\{\sigma\right\}
\label{mom1}
\end{equation}
where $\left[Z\right]$ is the square symmetric matrix built with $Z_{mn}$ as elements and the vector $\left\{\sigma \right\}$ is the list of charge densities for each triangle in which the surfaces of the conductors have been discretized. 
$\left[S\right]$ is a matrix with on the diagonal all the areas of the mesh elements:
\begin{equation}
	[S] = \begin{bmatrix} S_{1} & & 0 \\
	& \ddots & \\
	0 & & S_{N_A + N_B}
	\end{bmatrix}
\label{cmtot}
\end{equation}

In the application here dealt with, the voltages and the geometries of the two bodies are assigned and the charge densities are unknown.

The terms of matrix $\left[Z\right]$ are grouped in three types:
\begin{enumerate}
	\item \textbf{Far terms}: these are the terms where $n$ and $m$ are elements far from each other. That means that:
		\begin{equation}
	\left| \textbf{r}_n - \textbf{r}_m \right| >> \max{(l_1...l_6)}
\label{far0}
\end{equation}
	where $\textbf{r}_n$ and $\textbf{r}_m$ are the position of the center of the two elements respectively and $l_1...l_6$ are the lengths of their edges. The integration of Eq.~\ref{volt3} is then approximated by:
	\begin{equation}
	Z_{mn} = \frac{S_n S_m}{4\pi\epsilon_0 \left| \textbf{r}_n - \textbf{r}_m \right|}
\label{far}
\end{equation}
with $S_n$ and $S_m$ as areas of the two elements and $\left| \textbf{r}_n - \textbf{r}_m \right|$ the distance between the centers of the two elements.
		\item \textbf{Self terms}: these are the terms where $n = m$. The integration of $Z_{mn}$ is complex, but a closed-form solution is available \cite{MOM2,trimesh} and implemented.
		\item \textbf{Near terms}: these are the terms that do not fall in the previous categories. There is no closed-form solution for the integration of $Z_{mn}$. One simple option is to subdivide the elements in smaller ones such that the condition of Eq.~\ref{far0} applies. The relevant matrix element is then the sum of these sub-contributions.
\end{enumerate}

Once the $\left[Z\right]$ matrix is built, the charge densities are obtained by solving the linear system of Eq.~\ref{mom1} with the voltage as boundary condition, which is often convenient because in conductors each body has a uniform potential.

Moreover, we can obtain the capacitance matrix from Eq.~\ref{mom1}, where the LHS contains the voltage on each mesh element and the RHS contains the list of charge densities $\left\{\sigma\right\}$:
\begin{equation}
	[C] = [S] [Z]^{-1} [S]
\label{capacitance}
\end{equation}
By clustering the matrix elements with respect to the object they belong to and summing them, one obtains the capacitance matrix between the physical objects under analysis. In the case of Fig.~\ref{fig1} it is a $2 \times 2$:
\begin{equation}
	[C_{obj}] = \begin{bmatrix} \sum\limits_{i = 1}^{N_A} \sum\limits_{j = 1}^{N_A} C_{ij} & \sum\limits_{i = 1}^{N_A} \sum\limits_{j = N_A}^{N_B} C_{ij} \\
	\sum\limits_{i = N_A}^{N_B} \sum\limits_{j = 1}^{N_A} C_{ij} & \sum\limits_{i = N_A}^{N_B} \sum\limits_{j = N_A}^{N_B} C_{ij}
	\end{bmatrix}
\label{cm}
\end{equation}
where the terms on the diagonal are the self-capacitances.

%Knowing the charge distribution, the electric field $\textbf{E}(\textbf{r})$ is given by:
%\begin{equation}
	%\textbf{E}(\textbf{r}) = \sum_{n=1}^{N_A+N_B} \frac{a_n S_n}{4\pi\epsilon_0 \left|\textbf{r} - \textbf{r}_n\right|^3}\left(\textbf{r} - \textbf{r}_n\right)
%\label{field1}
%\end{equation}
%assuming that the position $\textbf{r}$ is not in the vicinity of any charged surface. Otherwise, the mesh elements in the vicinity should be subdivided in smaller ones.

Knowing the charge distribution, the force between body A and B is given by:
\begin{equation}
	\textbf{f}_{A-B} = \sum_{n=1}^{N_A}\:\sum_{m=N_A+1}^{N_A+N_B} \frac{\sigma_n S_n \sigma_m S_m}{4\pi\epsilon_0 \left|\textbf{r}_n - \textbf{r}_m\right|^3}\left(\textbf{r}_n - \textbf{r}_m\right)
\label{force1}
\end{equation}
For the near-terms mentioned above, one should use a local numerical integration over the surfaces of the elements pair. The electric field is obtained in a similar way.

As alternative, the force can be derived from the capacitance matrix with the following formula \cite{Jackson}:
\begin{equation}
	f_{A-B,x_i} =  \frac{1}{2} \frac{\partial C_{A-B}}{\partial x_i} V_{A-B}^2
	\label{force00}
\end{equation}
where $C_{A-B}$ is the capacitance between body A and B, $V_{A-B}$ is their relative voltage and $x_i$ is a general axis x, y or z. The torque is given by an equivalent formula:
\begin{equation}
	\tau_{A-B,\theta_i} =  \frac{1}{2} \frac{\partial C_{A-B}}{\partial\theta_i} V_{A-B}^2
	\label{torque00}
\end{equation}
Using Eq.~\ref{force00} requires the numerical estimation of the partial derivative and therefore solving Eq.~\ref{capacitance} more than once.

%%%%%%%%%%%%%%%%%%%%%%%%%%%%%%%%%%%%%%%%%%%%%%%%%%%%%%%%%%%%%%%%%%%%%%%%%%%%%%%%%%%%%%%%%%%%%%%%%%%%%%%%%%%%%%%%%%%%%%%%%%%%%%%%%%%%%%%%%%%%%%%%%%%%%%%%%%%%%%%%
\section{The LISA Pathfinder Gravitational Reference Sensor and the release phase of the test mass}
\label{sec2}

\begin{figure}[h!]
\begin{center}
\includegraphics[width=0.75\columnwidth]{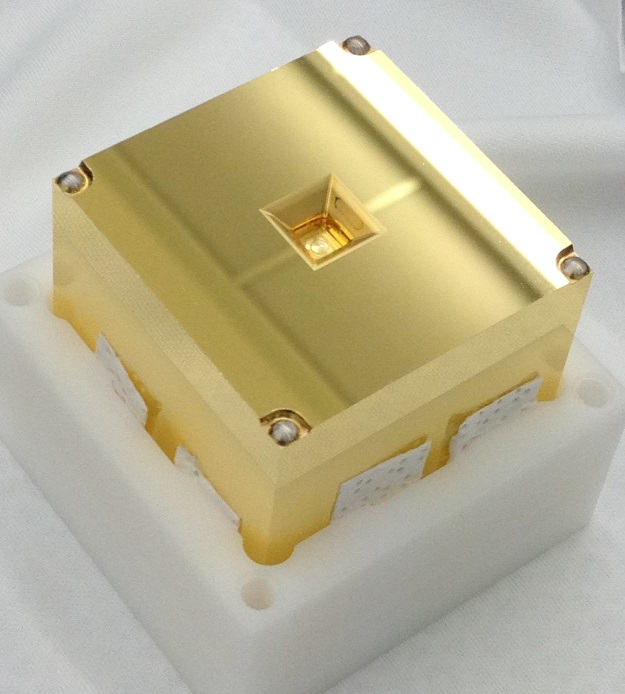}
\caption{View of a TM. The hole in the middle of the visible surface is one of the indents where the plungers are engaged. The hemispherical surfaces at the corners are the interfaces to the CVM. Image from \cite{TMimg}.}
\label{fig2}
\end{center}
\end{figure}

\begin{figure}[h!]
\begin{center}
\includegraphics[width=0.8\columnwidth]{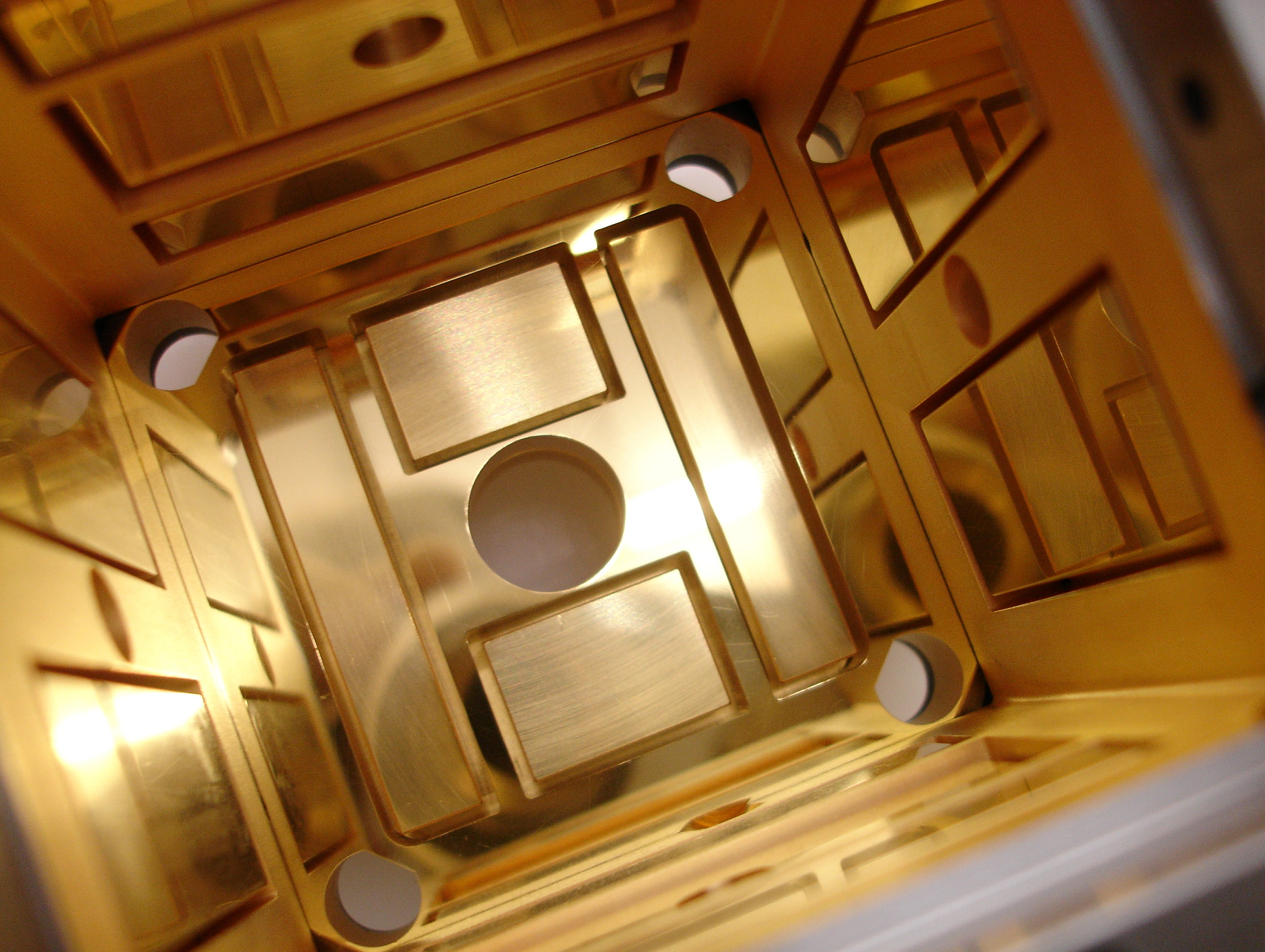}
\caption{View of a EH. The rectangular surfaces are the electrodes used for actuation and sensing purposes. Image from \cite{EHimg}.}
\label{fig3}
\end{center}
\end{figure}

Each LPF Gravitational Reference Sensor (GRS), implemented by OHB Italy, is made up of a test mass (TM, Fig.~\ref{fig2}), an Electrode Housing (EH, Fig.~\ref{fig3}) that performs actuation and sensing, a front end electronics (FEE), a precise lock and release mechanism symmetrical on two TM sides, and a discharge system, which keeps the TM neutral.

The TM is a 1.93~\si{\kilo \gram} 46~\si{\milli \metre} cube made of an 73:27 Au-Pt alloy. 
The EH is a molybdenum box with an empty cubic space for the TM. The EH hosts a set of electrodes whose voltage can be controlled, allowing actuation and sensing of the TM configuration.
Both EH inner surfaces and TM are gold coated \cite{LPFcharge1}.

\begin{figure}[h]
\begin{center}
\includegraphics[width=0.85\columnwidth]{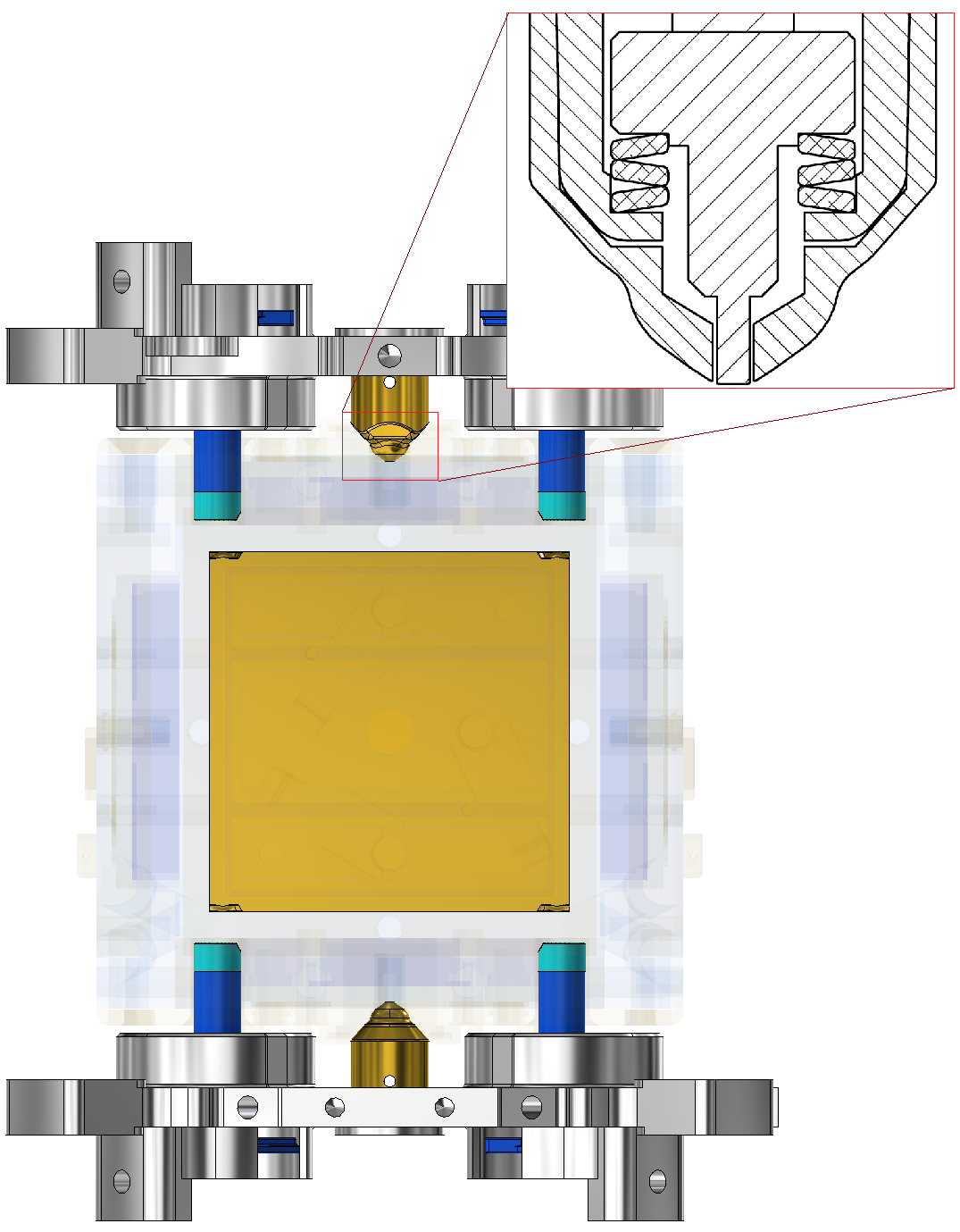}
\caption{CAD view of the end-effectors of CVM and GPRM, and TM. The CVM blue fingers engage the TM corners, clearly visible in Fig.~\ref{fig2}. The GPRM plungers are shown in gold color. The top-right image is an open view of the plunger head that shows the presence of the tip inside.}
\label{fig4}
\end{center}
\end{figure}

The lock and release mechanism is symmetrical on two TM sides and divided in two sub-systems, Fig.~\ref{fig4}, the Caging and Venting Mechanism (CVM) in charge of locking during launch \cite{Esmats2013z}, and the Grabbing Positioning and Release Mechanism (GPRM) that centers the TM and then leaves it free from any mechanical constraint \cite{ESMATS13i}. Both sub-systems have been developed by RUAG Space.

\begin{figure*}[!ht]
\begin{center}
\includegraphics[width=0.9\textwidth]{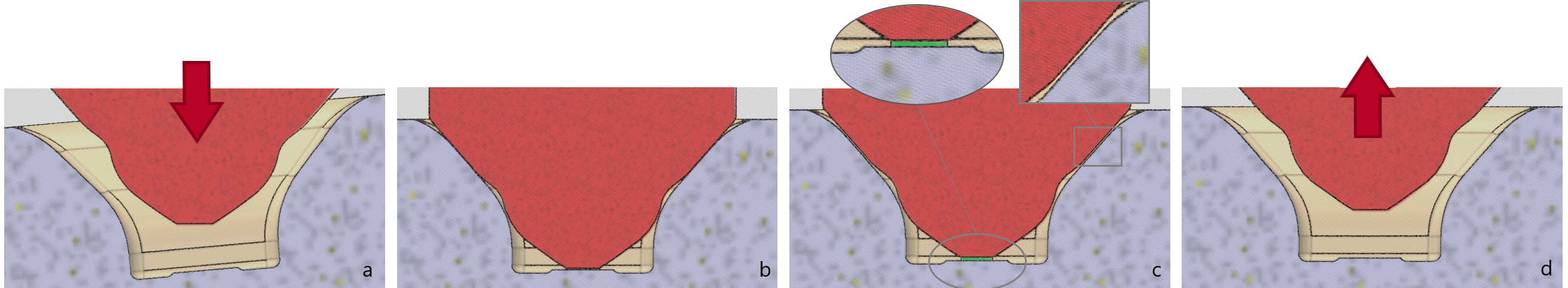}
\caption{Schematic representation of the mechanical part of the release procedure. An open view of a Plunger is in red color, the piezostack-moved tip is in green and the TM is in cyan. Referring to the list in the text: a) represents step 2 where the TM is off-centered and rotated by the CVM, b) step 3, c) step 4 d) step 5 and 6.}
\label{fig5}
\end{center}
\end{figure*}

The GPRM acts on the +z and -z TM faces where there are two pyramidal indents. The GPRM +z and -z side are slightly different as one of them has a pyramidal end-effector (or plunger) and the other one has an axial-symmetrical one, called cylindrical plunger. The tip of both plungers is actuated independently by means of a piezo-stack with a nominal extension of 15~\si{\micro \metre}, Fig.~\ref{fig4}. The function of the movable tips is to minimize the residual contact surface before the TM is detached from the GPRM, \cite{Benedetti2006}.

The nominal release procedure is qualitatively described by the following steps:
\begin{enumerate}
	\item the TM is fully constrained by the CVM eight fingers;
	\item the CVM fingers are detached from the TM, while the plungers move towards the TM, which in that instant can be off-centered and rotated due to the CVM effect;
	\item the plungers center the TM;
	\item a combined plunger retraction and tip extraction of about 15~\si{\micro \metre} ensures, nominally, that all the bonds created between plungers and TM are broken. Only a small residual contact surface at the tip end remains with a contact force of about 0.3~\si{\newton};
	\item t~=~0~\si{\second}. The tips are quickly retracted removing any residual bond by means of the TM inertia. 
	\item t~=~0+~\si{\second}. The plungers start moving. 
	\item t~=~10~\si{\second}. Activation of the Drag Free and Attitude Control System (DFACS, \cite{DFACS}) that performs electrostatic actuation in order to capture the TM and re-center it in the EH. It is worth noting that the TM state observer requires few seconds to have an accurate first estimate of the TM configuration;
	\item t~=~20~\si{\second}. The plungers are stopped after traveling for about 2.4~\si{\milli \metre};
	\item t~=~920~\si{\second}. If the TM is within the range of position, attitude and velocity prescribed for steady state, the plungers start the final retraction to their parking position outside the EH.
\end{enumerate}
Part of the release sequence is visualized schematically in Fig.~\ref{fig5}.

In the instant in which the last mechanical contact between TM and GPRM is broken, the TM becomes electrically charged due to the triboelectrification effect \cite{tribo}. This introduces electrostatic forces which can modify the performance of the release. In nominal operations, the charge management system uses UV light to remove excess charges coming from the initial triboelectrification condition and energetic cosmic and solar particles. Such a system is however not active during the release and capture phases and would be too slow \cite{LPFcharge1}. Even in LISA, where an UV-LED system is foreseen in place of the Hg lamps, the time constant is well above 1~\si{\s} \cite{UVLED}.

The requirements on the released-TM state are listed in Table \ref{tab1}.
\begin{table}[!b]
% increase table row spacing, adjust to taste
\renewcommand{\arraystretch}{1.3}
\caption{TM initial state requirement, relative to the EH.}
\label{tab1}
\centering
% Some packages, such as MDW tools, offer better commands for making tables
% than the plain LaTeX2e tabular which is used here.
\begin{tabular}{|c||c|}
\hline
\textbf{State} & \textbf{Value} \\
\hline
offset along x, y and z & $\pm$ 200~\si{\micro \metre} \\
\hline
linear velocity along x, y and z & $\pm$ 5~\si{\micro \metre / \second} \\
\hline
angle around x, y and z & $\pm$ 2~\si{\milli \radian} \\
\hline
angular rate around x, y and z & $\pm$ 100~\si{\micro \radian / \second} \\
\hline
\end{tabular}
\end{table}
Nominally, the DFACS is able to cope only with such initial state and an additional 50\% margin. 

Before flight, the linear velocity along z - the axis of the plungers motion - was considered the most challenging requirement because of the residual adhesion bonds \cite{Gane,ESASTM}. Adhesion is enhanced by its unpredictable behavior and the fact that all the mating surfaces are gold coated. Such a behavior is the reason why symmetry does not assure a net zero velocity \cite{MSSP13}. In the development phase, an anti-adhesive layer was considered on the tip, but eventually discarded to minimize impacts on the force noise budget and in light of the on-ground adhesion test results \cite{Bortoluzzi2004,ASR13,Trans,Bortoluzzi2017}.

In-orbit the actual results were far beyond the requirements, with peak values after the first release of 56.7~\si{\micro \metre / \second} and 1084~\si{\micro \radian / \second} \cite{esmats19}. The velocities encountered were not only along z, as was expected if adhesion was the main source of momentum. Nonetheless, thanks to a manual procedure and to the superior performance of the DFACS, it was possible to capture the TM and control it to steady-state. The manual release procedure consisted in a small retraction of the plungers, which still produced non compliant TM states but within a confined space around the EH center. The control was activated manually when the TM crossed the center of the housing \cite{Bortoluzzi2020}.

Later analysis highlighted that during a release with the nominal procedure the TM bounced in the EH and on the plungers before being captured. However, such a procedure is not reliable in terms of repeatability and integrity of the exposed surfaces. TM bounces on the plungers were eventually employed by the manual procedure. In view of LISA, where the manual procedure is not an option, the roles played by different phenomena must be clarified.

Qualitatively, the most probable reasons for the high velocities are:
\begin{enumerate}
	\item a combination of mechanical effects such as:
		\begin{itemize}
				\item non nominal contact configuration between TM and plungers (e.g. not only the tips are in contact);
				\item lack of tips' and plungers' synchronization in the release (i.e. the contact load is removed on one side before the other);
				\item spurious oscillations of the plungers that collide with the TM.
		\end{itemize}
	\item an underestimation of electrostatic forces.
\end{enumerate}
The second point is covered by this paper. For the mechanical effects, the investigation is presented in \cite{Bortoluzzi2020}.

The TM-to-housing voltage, when the mechanical steady state is achieved, has been measured in-flight and indicates the presence of electrostatic charges that contribute to the force acting on the TM.

%%%%%%%%%%%%%%%%%%%%%%%%%%%%%%%%%%%%%%%%%%%%%%%%%%%%%%%%%%%%%%%%%%%%%%%%%%%%%%%%%%%%%%%%%%%%%%%%%%%%%%%%%%%%%%%%%%%%%%%%%%%%%%%%%%%%%%%%%%%%%%%%%%%%%%%%%%%%%%%%
\section{Estimation of electrostatic effects on the release performance}

During the release and capture sequence there are two main electrostatic effects that influence the performance:
\begin{enumerate}
	\item the Coulomb force on the TM, mostly due to the charge accumulation and the very small gap of the plungers to the TM;
	\item the non nominal transient capacity measured by the electrodes: the presence of the two non identical plungers in motion inside the EH introduces a bias in the measurement.
\end{enumerate}

The voltage of TM, EH and GPRM right after the release is not known. It is extrapolated from this relationship, based on the charge conservation:
\begin{equation}
	C_{TM,t=0} V_{TM,t=0} = C_{TM,t=ss} V_{TM,t=ss}
	\label{charge0}
\end{equation}
where $t = ss$ is any time instant after t~=~920~\si{\second} or step 9 of the sequence introduced above. The capacitance $C_{TM}$ is the total capacitance between the TM and any other object. All the voltages are referred to the spacecraft. 

The total TM capacitance at t~=~$0^+$~\si{s}, $C_{TM,t=0}$, is 82.4~\si{\pico \farad}, if the tip length is 15~\si{\micro \metre} and TM and plungers are aligned. Using the MoM described in Section~\ref{sec1}, at steady state $C_{TM,t=ss}$ is 36.7~\si{\pico \farad}. A FE model mentioned in \cite{LPFcharge2} estimates that $C_{TM,t=ss}$ is 34.2~\si{\pico \farad}. The FE model is not documented except for the operational case when the plungers are outside the EH \cite{Nico}. 

The TM potential at steady state, $V_{TM,t=ss}$, is known from \cite{LPFcharge1}, which reports -421~\si{\milli \volt} and -281~\si{\milli \volt} for the first release of the two TM. Later releases showed potentials between -100~\si{\milli \volt} and -50~\si{\milli \volt}. With Eq.~\ref{charge0}, the -421~\si{\milli \volt} translates into a $V_{TM,t=0}$ of -188~\si{\milli \volt}, right after a nominal release. Such a figure is also consistent with the results of \cite{tribo} for an Au-Au contact, where the voltage difference is a normal distribution with zero mean and $2\times\sigma$ about 1~\si{\volt}.

\subsection{Electrostatic force between TM and plungers}

The force on the TM at any time is given by the following relationship:
\begin{equation}
	f_{TM,x_i}(t) =  \frac{1}{2} \frac{\partial C_{TM}}{\partial x_i} V_{TM,t}^2 =  \frac{1}{2} \frac{\partial C_{TM}}{\partial x_i} \left(\frac{C_{TM,t=ss} V_{TM,t=ss}}{C_{TM,t}}\right)^2
	\label{force0}
\end{equation}
where $x_i$ is a general axis x, y or z. The torque is given by an equivalent formula, see Eq.~\ref{torque00}.

%Considering that the plungers are at about 15~\si{\micro \meter} distance while all the electrodes are at a few~\si{mm}, one can assume that $\frac{\partial C_{\text{TM}}}{\partial q} \approx \frac{\partial C_{\text{TM-plunger}}}{\partial q}$.

The evaluation of the electrostatic contribution to the release is done via a limit-case approach in which the first instants of the release are analysed for scenarios considered at the boundary of the space of the possible ones.

\begin{figure}[h!]
\begin{center}
\includegraphics[width=\columnwidth]{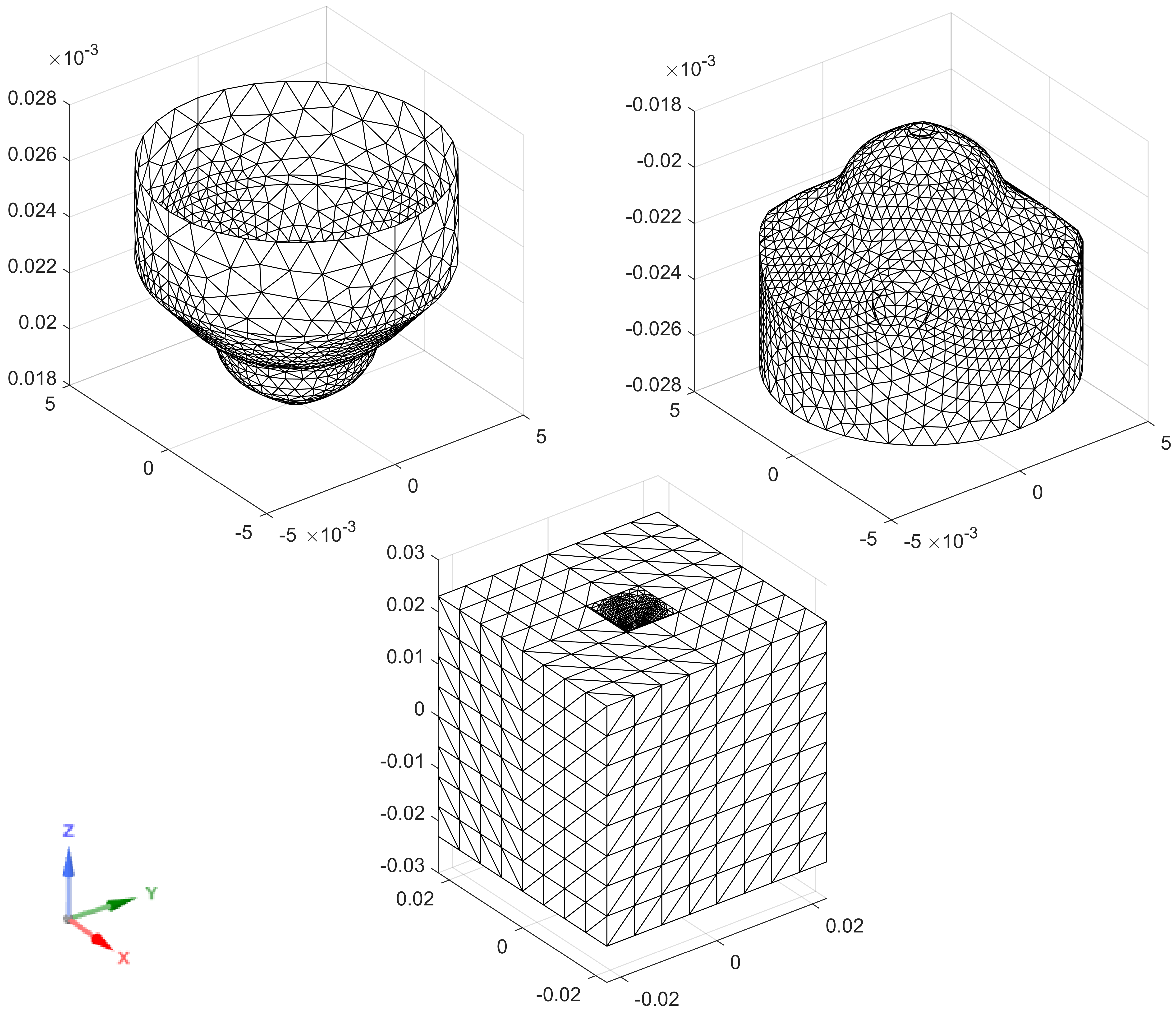}
\caption{View of the mesh of the TM, pyramidal plunger and cylindrical plunger surfaces. Lengths are in [\si{\meter}].}
\label{fig6}
\end{center}
\end{figure}
\begin{figure}[h!]
\begin{center}
\includegraphics[width=0.7\columnwidth]{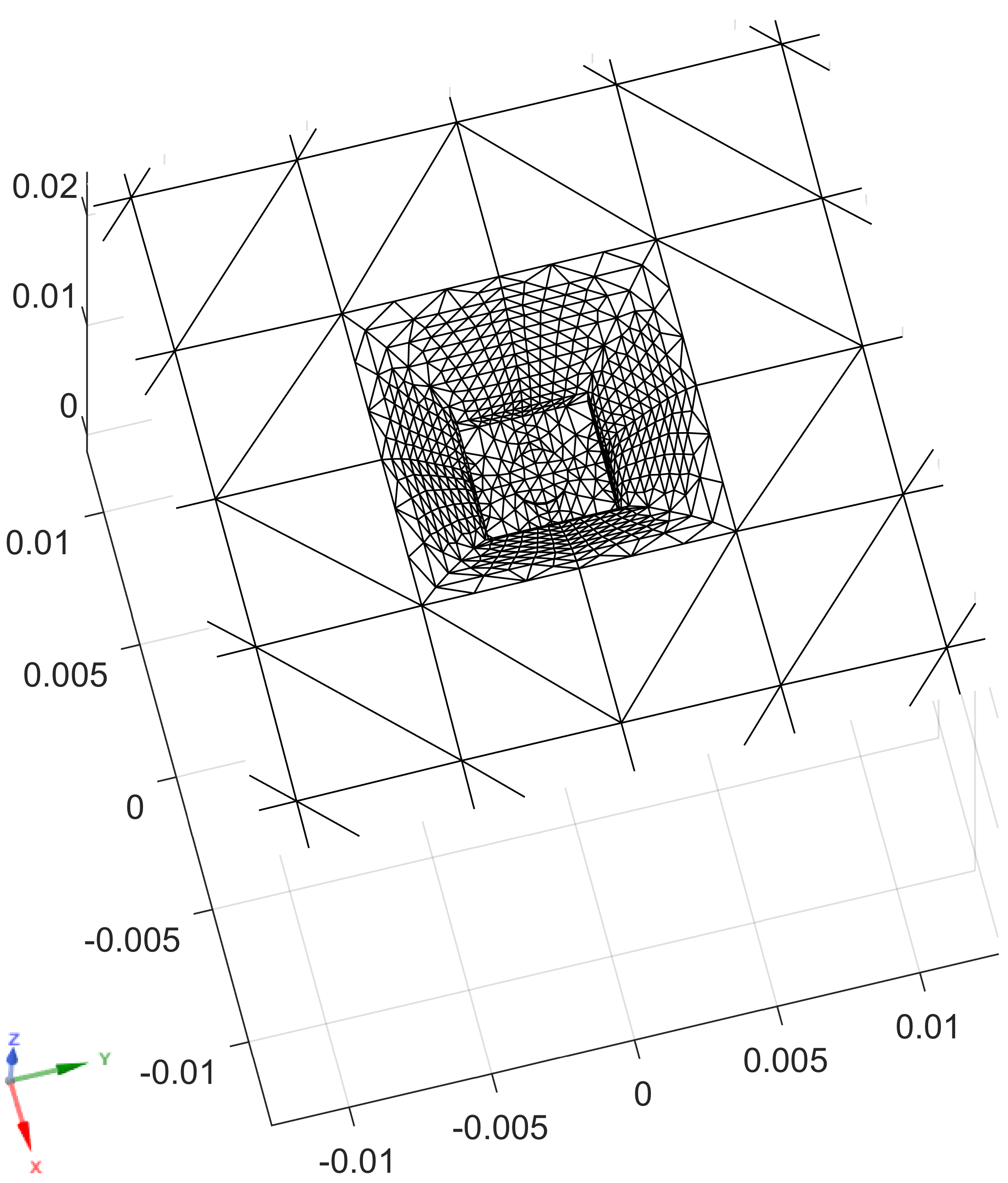}
\caption{Zoomed view on the mesh of the TM indent. Lengths are in [\si{\meter}].}
\label{fig7}
\end{center}
\end{figure}

The entire EH, TM and plunger surfaces are included in the simulation. The triangular mesh generated on their surfaces is shown in Fig. \ref{fig6} and \ref{fig7}. Small fillets at the TM indents edges and the CVM mating surfaces have been removed to limit the number of mesh elements. The removed surfaces are all far from both plungers and EH. 

The coordinate system is centered on the TM and has the z-axis aligned with the direction of the release. The x and y-axis are indistinguishable for what concerns this work.

The threshold between far-terms and near-terms is set to 650~\si{\micro \meter}, which is a number sufficiently high to cover well the elements of plunger and TM in proximity, but low enough to avoid integrating $Z_{mn}$ of Eq.~\ref{volt3} on too many mesh triangles.

The limit-case scenarios differ from the nominal configuration (perfect alignment, initial gap of 15~\si{\micro \metre} and TM in the center of the EH without any momentum) for the following:
\begin{enumerate}
	\item case \#1: the pyramidal plunger in t = 0~\si{s} is retracted from the TM by 10~\si{\micro \metre}, the cylindrical plunger by 25~\si{\micro \meter}. This accounts for a different extension of the tip piezo-stacks.
	\item case \#2: at t = 0~\si{s}  the TM is displaced of 5~\si{\micro \meter} along the y-axis;
	\item case \#3: the TM is rotated of 0.2~\si{\milli \radian} about the y-axis;
	\item case \#4: the TM is rotated of 3~\si{\milli \radian} about the z-axis;
	\item case \#5: the pyramidal plunger in t = 0~\si{s} is at a distance of 10~\si{\micro \metre}, the TM is rotated of 0.05~\si{\milli \radian} about the y-axis and 1~\si{\milli \radian} about the z-axis.
	\item case \#6: like case \#1, but with the pyramidal plunger stopping at 100~\si{\micro \metre}.
\end{enumerate}
The plungers are assumed starting their motion 0.1~\si{s} after the tip quick movement starts.

\begin{table*}[tb!]
%\tiny
\renewcommand{\arraystretch}{1.3}
\caption{Peak force/torque and change in TM velocity}
\label{tab2}
\centering
\begin{tabular}{l||l|l|l|l|l|l||l|l|l|l|l|l|}
\cline{2-13}
                             & \multicolumn{6}{c||}{\textbf{Magnitude of Peak Force or Torque}} & \multicolumn{6}{c|}{\textbf{$\Delta v$ or $\Delta \omega$ TM }} \\ 
\cline{2-13}
														 & \multicolumn{3}{c|}{[\si{\micro \newton}]} & \multicolumn{3}{c||}{[\si{\micro \newton\metre}]} & \multicolumn{3}{c|}{[\si{\micro \metre/\second}]} & \multicolumn{3}{c|}{[\si{\micro \radian/\second}]} \\ \hline
\multicolumn{1}{|l||}{}       	& x    		& y    		& z    		& x    	& y    		& z   			& x   			& y   		& z  				& x  		& y  		& z  			\\ \hline
\multicolumn{1}{|l||}{Nominal} 	& -    		& -    		& $0.4$   & -    	& -    		& -   			& -   			& -   		& $-0.06$  	& -  		&-   		& -  			\\ \hline
\multicolumn{1}{|l||}{Case \#1} & -    		& -    		& $1.3$   & -    	& -    		& -   			& -   			& -   		& $-0.14$  	& -  		&-   		& -  			\\ \hline
\multicolumn{1}{|l||}{Case \#2} & -    		& $0.6$   & $0.8$   & $0.02$& -    		& -   			& -         &$0.05$		& $-0.09$    &$-3.8$ & -   	& -   		\\ \hline
\multicolumn{1}{|l||}{Case \#3} & $0.7$   & -    		& $0.9$   & -   	& $0.02$  & -   			& $0.05$    & -   		& $-0.09$  	& -  		& $4.1$ & -   		\\ \hline
\multicolumn{1}{|l||}{Case \#4} & -    		& -    		& $6.9$   & -    	& -    		& -       	& -   			& -   		& $-0.08$  	& -  		& -   	& - 			\\ \hline
\multicolumn{1}{|l||}{Case \#5} & -       & -    		& $0.5$   & -    	& -     	&  -   			& -       	& -   		& $-0.06$  	& -  		& -     & -				\\ \hline
\multicolumn{1}{|l||}{Case \#6} & -       & -    		& $1.3$   & -    	& -     	&  -   			& -       	& -   		& $-0.21$  	& -  		& -     & -				\\ \hline
\end{tabular}
\end{table*}

The numerical problem is coded and solved in MATLAB~R2019b. The analyzed time is 10~\si{\second} which is sufficient to move the plungers far from the TM, i.e. more than 1~\si{\milli \metre}.

For simplicity, the TM position is assumed not to change in the time considered. This assumption is justified by the results explained below, where the speed gained by the TM is far below the plungers retraction velocity of 120~\si{\micro \metre/\second}. It is worth noting that the purpose of this analysis is not to find the electrostatic effect when the TM has a high velocity, but to find if the electrostatic contribution can alone justify the high values experienced in-flight.

At each time-step, the force and torque are given by solving the MoM problem with a unitary voltage on the TM and then scaling it according to Eq.~\ref{force0}, with the capacitance, $C_{\text{TM,t}}$, estimated at the same time-step. The steady state voltage is assumed in all the simulations to be twice the worst-case experienced in-flight, i.e. 421~\si{\milli \volt}. We also add 200~\si{\milli \volt} at all times to cover local patch effects \cite{Buchman2011}. %The 200~\si{\milli \volt} \hl{are applied on all the TM surface. In the limit-case approach we follow, that is equivalent to having patch effects always where the gap is small. Potentials on other surfaces do not } %\hl{This is a conservative assumption because after the releases performed during the science phase the voltage ranged between} -98~\si{\milli \volt} and -52~\si{\milli \volt}.

The changes in velocity and rate ($\Delta v$ and $\Delta \omega$) are obtained via the classical:
\begin{equation}
	m_{TM} \Delta \textbf{v} =  \int_{t_0}^{t_1} \textbf{f}_{TM}(t) dt
	\label{force2}
\end{equation}
and
\begin{equation}
	I_{TM} \Delta \bm{\omega} =  \int_{t_0}^{t_1} \bm{\tau}_{TM}(t) dt
	\label{torque2}
\end{equation}
where $m_{TM}$ and $I_{TM}$ are the TM mass and inertia, and $\textbf{f}_{TM}(t)$ and $\bm{\tau}_{TM}(t)$ are the force and torque acting on it, with $t_0$ = 0~\si{\second} and $t_1$ = 10~\si{\second}. Eq. \ref{torque2} is derived from the Euler's rotation equations assuming that angles and angular velocities are small.

\begin{figure}[!h]
\begin{center}
\includegraphics[width=0.9\columnwidth]{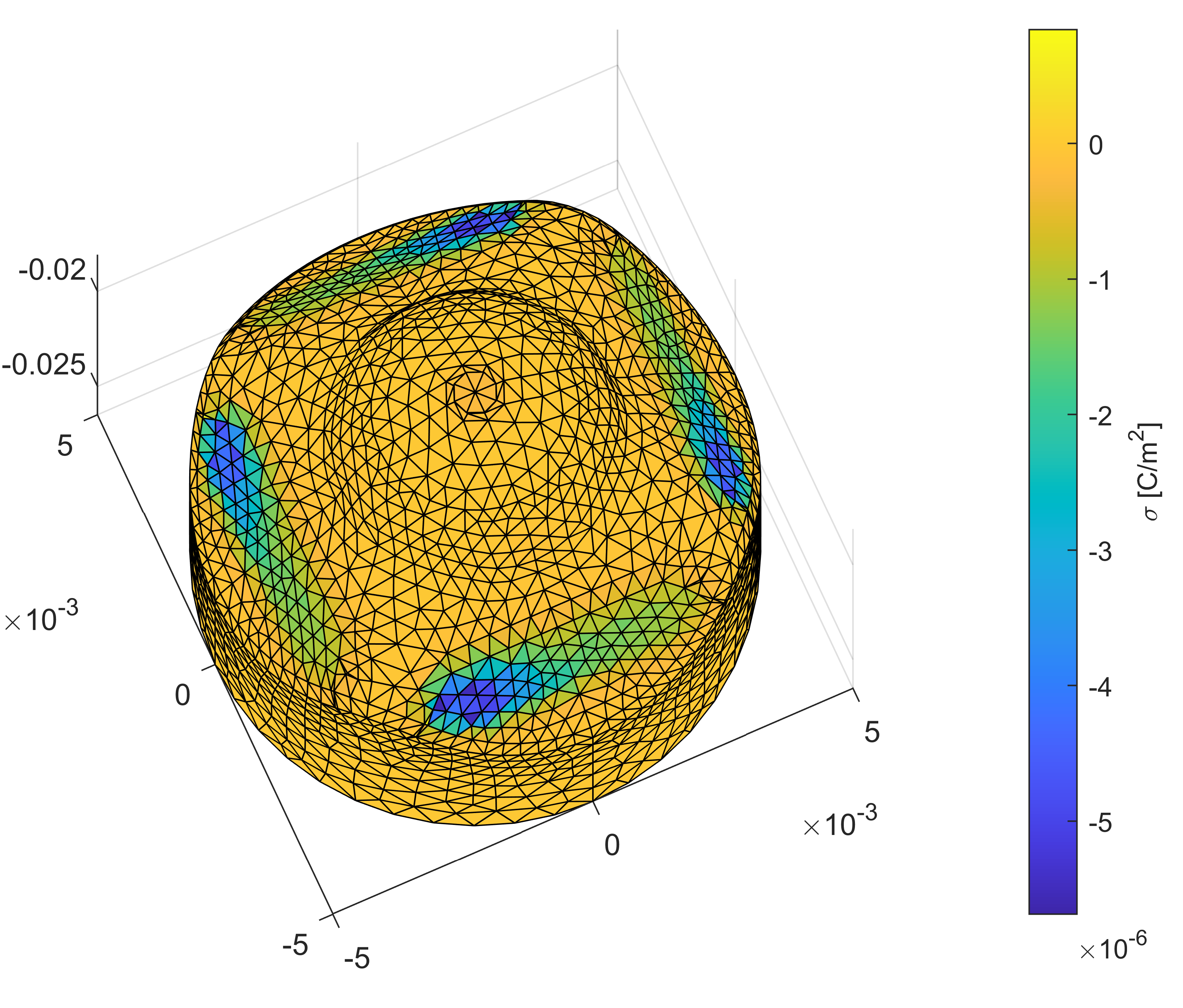}
\caption{Charge density on the pyramidal plunger surface in case \#4 and with a voltage on the TM of 1 V. Lengths are in [\si{\meter}].}
\label{fig8}
\end{center}
\end{figure}

\begin{figure}[!h]
\begin{center}
\includegraphics[width=0.9\columnwidth]{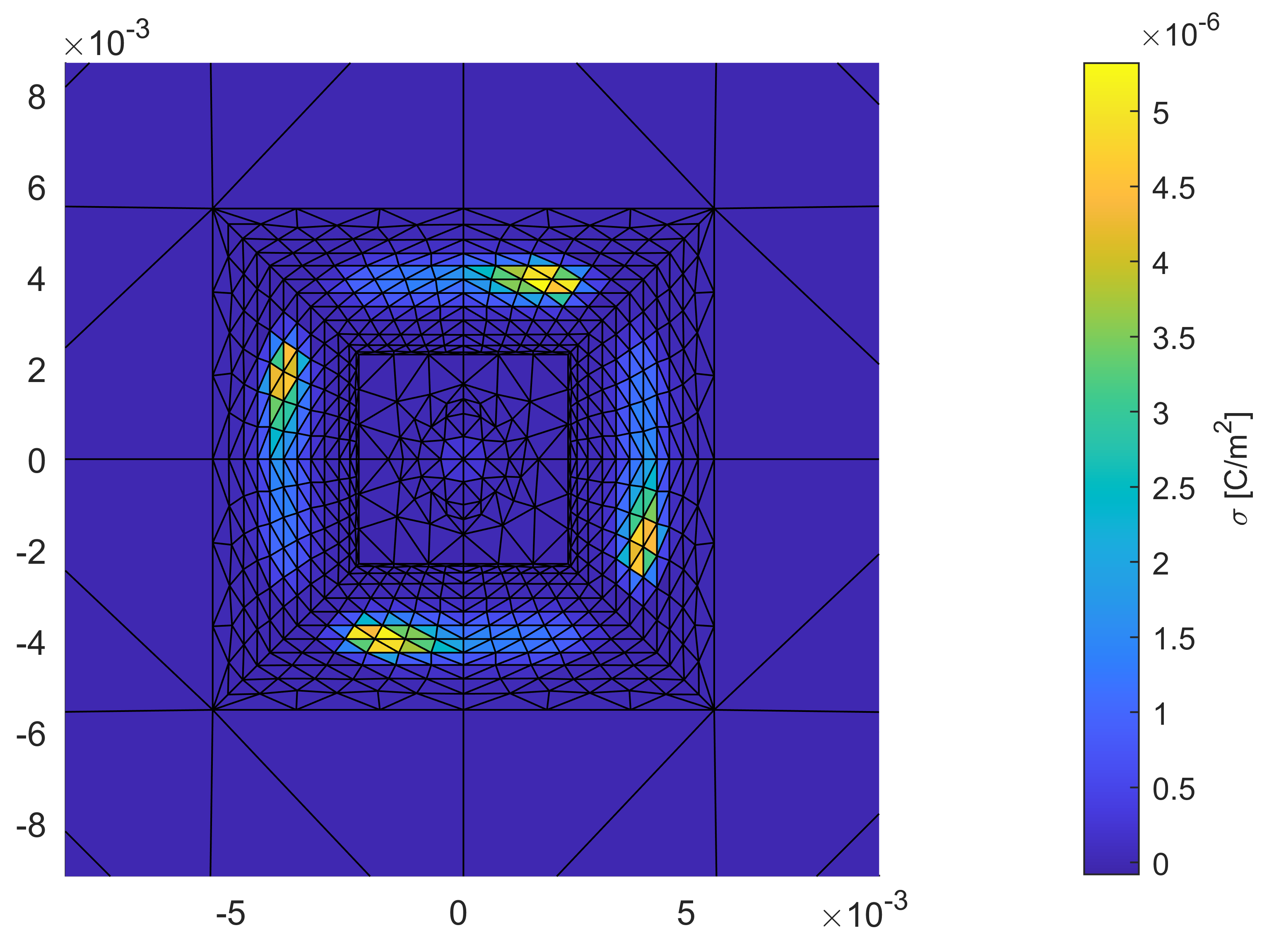}
\caption{Charge density on the TM surface in case \#4 and with a voltage on the TM of 1~\si{\volt}. Lengths are in [\si{\meter}].}
\label{fig9}
\end{center}
\end{figure}

The results obtained in the cases listed above are reported in Table \ref{tab2}, where the nominal scenario is also included for benchmarking reasons. An example of charge density is shown in Fig.~\ref{fig8} and \ref{fig9}. In the table, values below $10^{-2}$ are not reported. 

In the limit-case scenarios analysed, the TM velocities due to the electrostatic attraction from the plungers are always low even in comparison with the estimated adhesion effect \cite{Trans,ASR13}. A key parameter in the results is the time at which the plungers start moving. From that moment on, the force is reduced very quickly. Our assumption of 0.1~\si{s} is conservative. Case \#6 suggests that the plungers do not need to be retracted far to limit the Coulomb attraction.


To prove the statements above even in an extreme-case, we configure TM and plungers in such a way that only one of the plungers is attracting the TM and there is a minimum gap over the maximum possible area. That means that the pyramidal plunger is retracted of only 0.67~\si{\micro \metre} and the cylindrical plunger is considered far. The 0.67~\si{\micro \metre} retraction is derived from a minimum gap given by the sum of the roughness of the plunger and TM mating surfaces, which are an N5 and a N3 (ISO grade numbers, \cite{ISO}). We can assume that, below this value, a contact is keeping the TM at the same voltage of the spacecraft. The gap is then projected on the z-axis, i.e. divided by the cosine of the inclination angle of the surfaces (\ang{48.5}), to obtain the retraction. 

To model the electrostatic environment with such a small distance on a large area, we use a finer mesh than what is shown in Fig.~\ref{fig6} and \ref{fig7}. 

The result is a velocity of the TM along z of about 4.5~\si{\micro \metre/ \second}, still including the 0.1~\si{s} delay in moving the plungers. Despite being an extreme case, this result does not justify what has been experienced in-flight, which is bigger and has high velocities also in x, y and rotation degrees of freedom.

\subsection{Ghost velocity due to the plungers shape and motion asymmetry}

The second electrostatic effect we analyze is due to the presence of the non identical plungers in motion inside the EH, which is a source of unbalanced capacitance as seen by the electrodes. The reasons of such effect are:
\begin{itemize}
	\item different shapes of the plungers;
	\item asymmetrical motion of the plungers. The plungers are driven by a PI\textregistered Nexline  actuator commanded in steps \cite{ESMATS13i}, with a nominal retraction velocity of 120~\si{\micro \metre/\second} \cite{DFACS}. Conservatively, the range of retraction velocities is assumed between 100~\si{\micro \metre/\second} -- 140~\si{\micro \metre/\second}.
\end{itemize}

\begin{figure}[b!]
\begin{center}
\includegraphics[width=0.7\columnwidth]{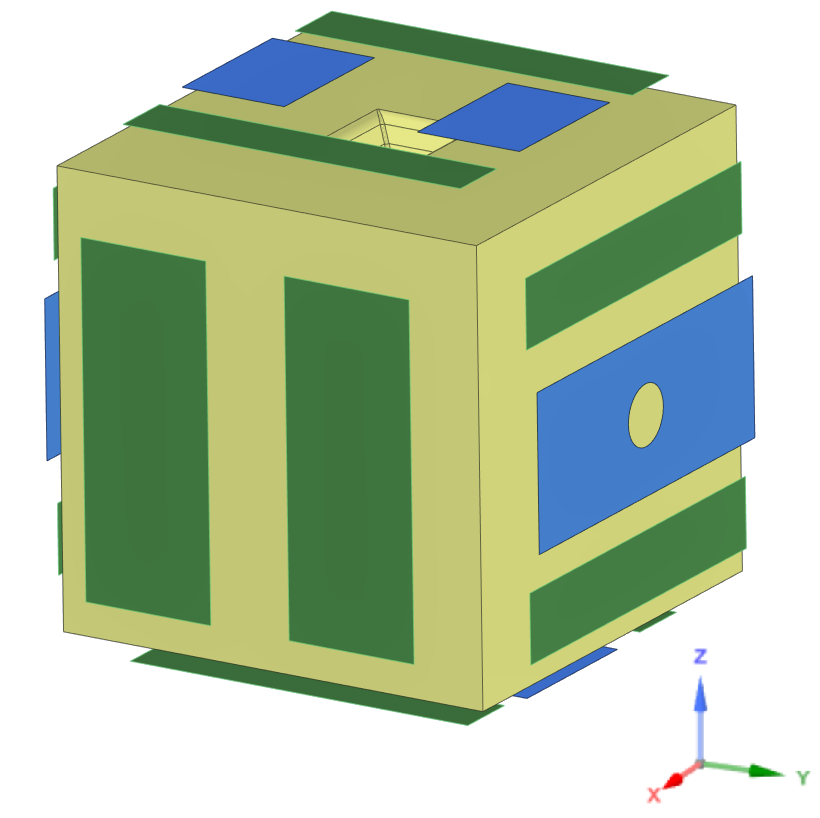}
\caption{Configuration of the electrodes surrounding the TM. The blue electrodes are used to induce a 100~\si{\kilo \hertz} voltage on the TM, which allows the sensing function via the green electrodes whose capacitance to the TM is measured.}
\label{fig10}
\end{center}
\end{figure}

The capacitance between the TM and the electrodes is used for sensing purposes. Considering only the linear example along z, the reference zero position of the TM is when the capacitance measured between the TM and the +z electrodes equals the one between TM and -z electrodes., in the electrodes configuration of Figure \ref{fig10}.

As a consequence of the unbalanced capacity that varies with time, the electronics sees a ghost motion of the TM. Such an effect is estimated by modeling the EH capacitive matrix and extracting the total capacitance of the electrodes in charge of measuring the motion along z. The unbalanced capacitance due to the plungers is equivalent to having the TM in a position that differs from its real one.

Assuming parallel plate capacitors, the change in capacitance as a function of a TM motion is:
\begin{equation}
	\Delta C \approx  C_{TM-EHz,0} \frac{d_z-z}{d_z}-C_{TM-EHz,0} \frac{d_z+z}{d_z}
	\label{dcap}
\end{equation}
where $C_{TM-EHz}$ is the capacitance between the z-sensing electrodes and the centered TM and $d_z$ is the gap in the $z$ direction. If the TM is not moving, the ghost position is:
\begin{equation}
	z(t) = -\frac{\Delta C(t) \: d_z}{2 C_{TM-EHz,0}}
	\label{dcap1}
\end{equation}

\begin{figure}[h!]
\begin{center}
\includegraphics[width=\columnwidth]{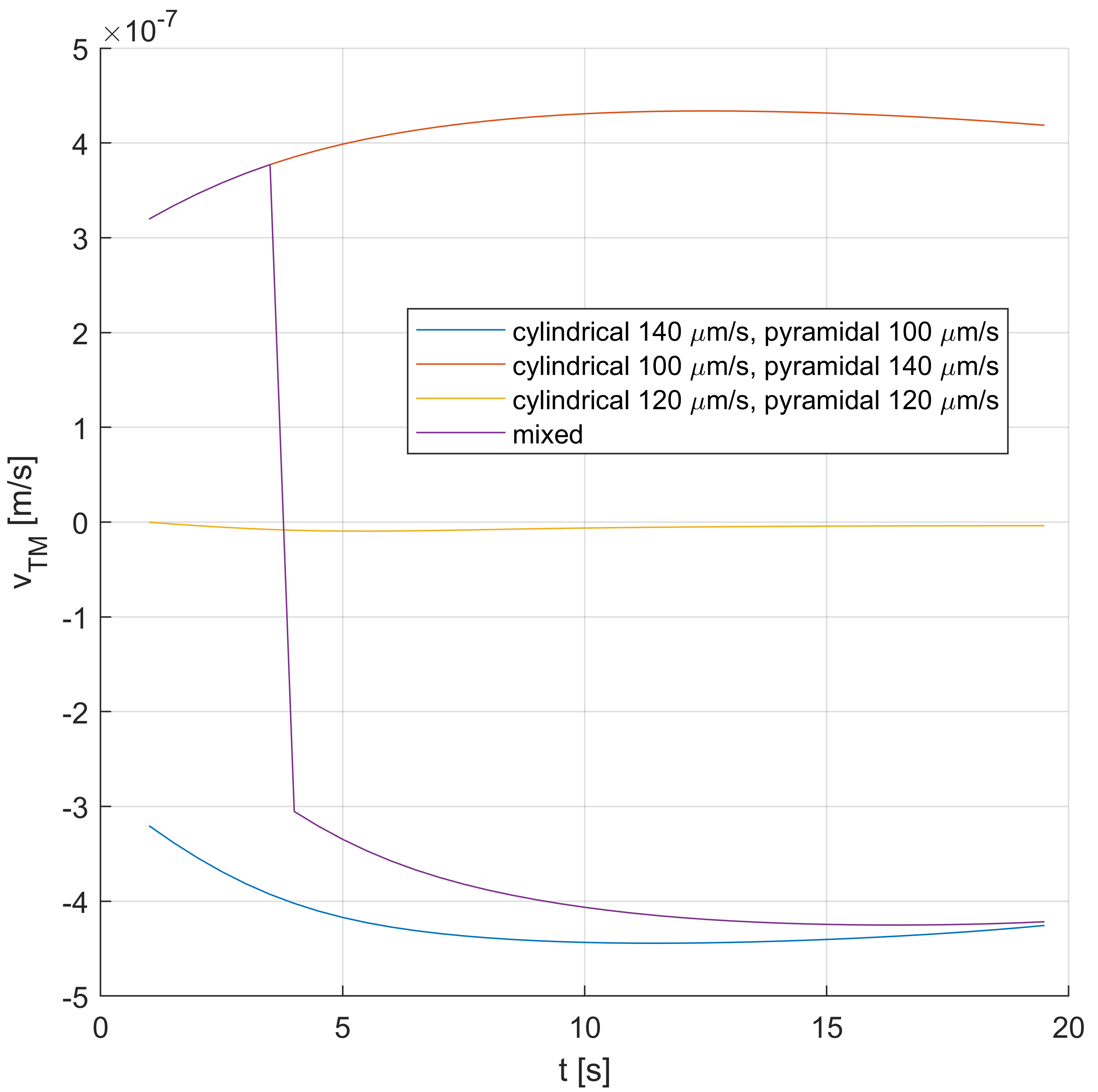}
\caption{Ghost TM velocity seen by the capacitive sensors due to a non symmetrical motion of the plungers. The "mixed" curve refers to a case in which the pyramidal plunger moves at 140~\si{\micro \metre/\second} in the first 4~\si{s} and at 100~\si{\micro \metre/\second} afterward, with the cylindrical plunger doing the opposite.}
\label{fig11}
\end{center}
\end{figure}

The first derivative of this position in time is the ghost velocity, Fig.~\ref{fig11}. $\Delta C$ is estimated with the MoM. A $\pm$~20~\si{\micro \metre/\second} difference on the nominal plunger velocity of 120~\si{\micro \metre/\second} determines 0.45~\si{\micro \metre/\second} of TM ghost velocity. The scenario in which the plungers retraction speeds change in time can give similar outcomes, Fig.~\ref{fig11}, but requires modeling all the sensing electronics to be understood accurately.

The magnitude of the effect of Figure~\ref{fig11} is limited and does not provide an explanation for the non-compliant in-flight releases. The plungers shape asymmetry alone, when the retraction speed is equal, has a low impact.

\subsection{Design considerations}

The electrostatic contribution is about a factor 2 lower than what has been estimated on-ground as adhesion effect and cannot justify the in-flight non-compliance. However, it can give a few indications on how to improve the release.

For LISA, the following shall be considered:
\begin{itemize}
	\item delay in the plungers retraction is key in the amount of momentum transferred and shall be controlled and minimized;
	\item there is no electrostatic reason to retract the plungers very far from the TM in non-science modes;
	\item shape tolerances and plungers motion should guarantee that the gap is always above few~\si{\micro \metre};
	\item the capacitive position sensing benefits from a smooth and symmetric retraction of the plungers.
\end{itemize}

\section{Conclusion}
LISA-Pathfinder has proven to be a robust and reliable machine. However, the dynamics and performance of the release phase, in which the test masses are set into free fall, were not in line with the formal requirement, the design and the on-ground expectations.

In view of LISA, the reasons of this non-compliance shall be understood and solved. We estimate the electrostatic contribution to be below 0.5~\si{\micro \metre/\second} and 5~\si{\micro \radian/\second}. Extreme, but unlikely, configurations can justify release velocities in the order of maximum few \si{\micro \metre/\second}. Coulomb forces and unbalanced sensor capacitances are therefore not the root cause of the behavior experienced in LISA-Pathfinder.

The estimate is based on a model of the electrostatic environment inside the Gravitational Reference Sensor built via the Method of Moments numerical technique, which is here over-viewed. Such a method and our model are useful not only for the release phase, but also for simulating the LISA test mass discharging system and the test mass control system.

% if have a single appendix:
%\appendix[Proof of the Zonklar Equations]
% or
%\appendix  % for no appendix heading
% do not use \section anymore after \appendix, only \section*
% is possibly needed

% use appendices with more than one appendix
% then use \section to start each appendix
% you must declare a \section before using any
% \subsection or using \label (\appendices by itself
% starts a section numbered zero.)
%

%\appendices
%\section{Proof of the First Zonklar Equation}
%Appendix one text goes here.

% you can choose not to have a title for an appendix
% if you want by leaving the argument blank
%\section{}
%Appendix two text goes here.

% use section* for acknowledgment
%\section*{Acknowledgment}
%The author would like to thank ...

%The authors would like to thank...

% Can use something like this to put references on a page
% by themselves when using endfloat and the captionsoff option.
\ifCLASSOPTIONcaptionsoff
  \newpage
\fi

% trigger a \newpage just before the given reference
% number - used to balance the columns on the last page
% adjust value as needed - may need to be readjusted if
% the document is modified later
%\IEEEtriggeratref{8}
% The "triggered" command can be changed if desired:
%\IEEEtriggercmd{\enlargethispage{-5in}}

% references section

% can use a bibliography generated by BibTeX as a .bbl file
% BibTeX documentation can be easily obtained at:
% http://mirror.ctan.org/biblio/bibtex/contrib/doc/
% The IEEEtran BibTeX style support page is at:
% http://www.michaelshell.org/tex/ieeetran/bibtex/
%\bibliographystyle{IEEEtran}
% argument is your BibTeX string definitions and bibliography database(s)
%\bibliography{IEEEabrv,../bib/paper}
%
% <OR> manually copy in the resultant .bbl file
% set second argument of \begin to the number of references
% (used to reserve space for the reference number labels box)
\bibliographystyle{IEEEtran}

% argument is your BibTeX string definitions and bibliography database(s)
%\bibliography{IEEEabrv,../bib/paper}
%
% <OR> manually copy in the resultant .bbl file
% set second argument of \begin to the number of references
% (used to reserve space for the reference number labels box)

% \begin{thebibliography}{1}

\bibliography{IEEEabrv,References} 

% biography section
% 
% If you have an EPS/PDF photo (graphicx package needed) extra braces are
% needed around the contents of the optional argument to biography to prevent
% the LaTeX parser from getting confused when it sees the complicated
% \includegraphics command within an optional argument. (You could create
% your own custom macro containing the \includegraphics command to make things
% simpler here.)
%\begin{IEEEbiography}[{\includegraphics[width=1in,height=1.25in,clip,keepaspectratio]{mshell}}]{Michael Shell}
% or if you just want to reserve a space for a photo:

\begin{IEEEbiography}[{\includegraphics[width=1in,height=1.25in,clip,keepaspectratio]{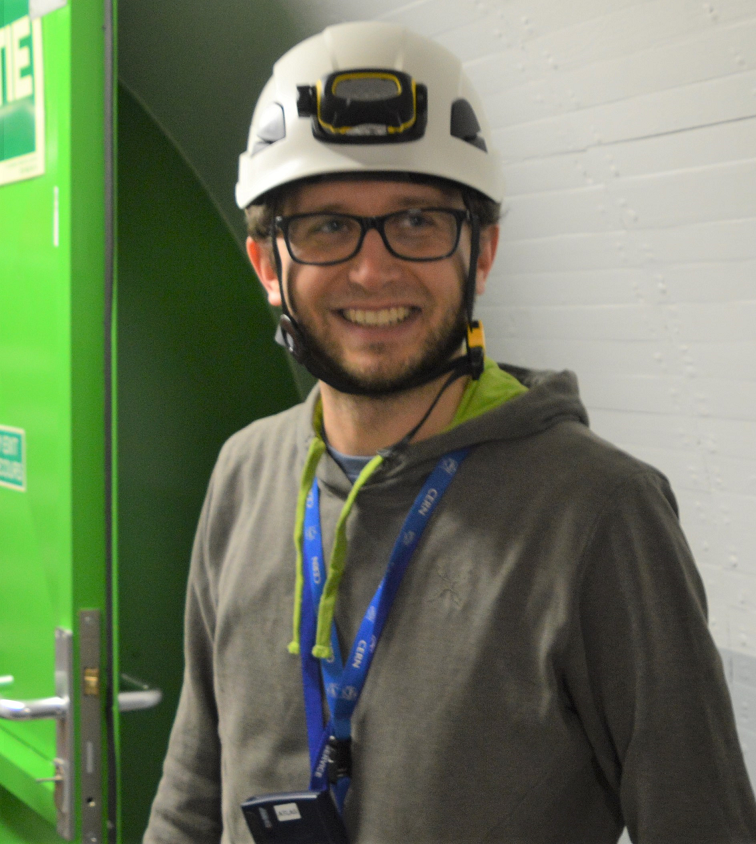}}]{Carlo Zanoni} has a Ph.D. in Mechanical Systems from the University of Trento (IT). He is now a Mechanical Engineer at the European Southern Observatory, based in Munich (DE). Before, he has worked at CERN (CH) from 2014 to 2017, on novel superconducting systems for the Large Hadron Collider, and at INFN (IT) from 2011 to 2014, on the LISA-Pathfinder European Space Agency mission, focusing on the release phase and on the development of future drag-free technologies. He  has also served as Visiting Researcher at Stanford University (USA), in 2012, and as Payload System Engineer in Airbus Space (formerly Astrium), between 2010 and 2011.
\end{IEEEbiography}

\begin{IEEEbiography}[{\includegraphics[width=1in,height=1.25in,clip,keepaspectratio]{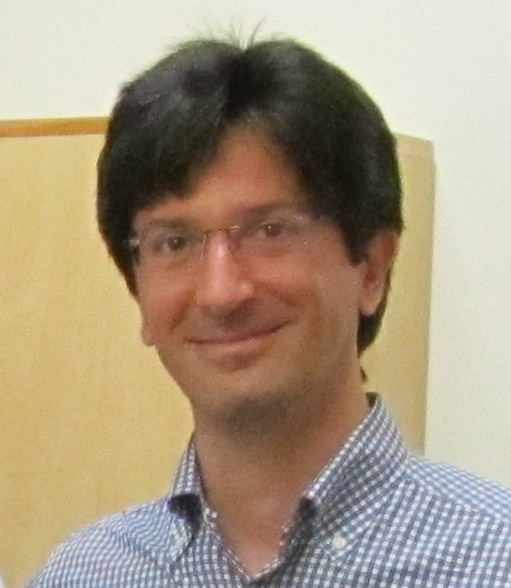}}]{Daniele Bortoluzzi} graduated in Mechanical Engineering and earned the Ph.D. in Mechanics of Machines at the University of Padova (IT) in the field of vehicle dynamics. He is Associate Professor of Mechanical Systems at the Department of Industrial Engineering of the University of Trento (IT). His research activity is mainly focused on the dynamics and control of mechatronics systems for space applications. He was part of the PI team of the LISA-Pathfinder mission, launched in late 2015, where he contributed to the design of the scientific payload architecture, spacecraft drag-free control and was responsible for the qualification of the release mechanism. He is now part of the team working on the development of the phase A of the joint ESA-NASA LISA mission. 
\end{IEEEbiography}

% if you will not have a photo at all:

% insert where needed to balance the two columns on the last page with
% biographies
%\newpage

% You can push biographies down or up by placing
% a \vfill before or after them. The appropriate
% use of \vfill depends on what kind of text is
% on the last page and whether or not the columns
% are being equalized.

%\vfill

% Can be used to pull up biographies so that the bottom of the last one
% is flush with the other column.
%\enlargethispage{-5in}

% that's all folks
\end{document}